\documentclass[a4paper]{spie}  

\usepackage{amsmath,amssymb}
\usepackage{graphicx}

\newcommand {\Ld} {{L_{\rm d}}}
\newcommand {\La} {{L_{\rm a}}}
\newcommand {\Lrad} {{L_{\rm r}}}

\newcommand {\bfJ}  {{\bf J}}

\newcommand {\dUmax} {U^{\prime}_{\max}}

\renewcommand {\d}  {{\rm d}}

\newcommand {\ee}  {{\rm e}}
\newcommand {\ch}  {{\rm ch}}

\newcommand {\E}  {{\varepsilon}}
\newcommand {\om}  {{\omega}}
\newcommand {\Om}  {{\Omega}}

\newcommand {\cD}  {{\cal D}}

\newcommand {\teta}  {{\tilde{\eta}}}

\newcommand {\kp}  {{\kappa}}

\newcommand {\amax}  {{a_{\max}}}
\newcommand {\pmax}  {{p_{\max}}}
\newcommand {\atf}   {{a_{\rm TF}}}

\title{Parameters of the crystalline undulator and its radiation for particular
experimental conditions}

\author{
A.~V.~Korol\supit{a}, A.~V.~Solov'yov\supit{a} 
and
W.~Greiner\supit{a,b}
\skiplinehalf
\supit{b} Frankfurt Institute for Advanced Studies,
          Johann Wolfgang Goethe-Universit\"at,
          Frankfurt am Main, Germany
\\
\supit{c} Institut f\"{u}r Theoretische Physik, 
          Johann Wolfgang Goethe-Universit\"at,
          Frankfurt am Main, Germany
}  
\authorinfo{
Further author information: 
A.V.K. is on leave from 
Department of Physics, State Maritime Technical University,
St.~Petersburg,  Russia.
A.V.S. is
on leave from Ioffe Physical-Technical Institute,
Russian Academy of Sciences, 
St.~Petersburg, Russia.}

\begin{document}
  \maketitle 

\begin{abstract}
We report the results of theoretical and numerical analysis of 
the crystalline undulators planned to be used 
in the experiments which are the part of the ongoing 
PECU project\cite{PECU}.  
The goal of such an analysis was to define the parameters (different 
from those pre-set by the experimental setup)  of the 
undulators which ensure the highest yield of photons of specified energies.
The calculations  were performed for 0.6 and 10 GeV positrons channeling
through periodically bent Si and Si$_{1-x}$Ge$_x$ crystals.
\end{abstract}


\section{INTRODUCTION} \label{Introduction}

In this paper we report the results of calculations 
of the parameters of the crystalline undulators 
(different from those pre-set by the experimental setup, see 
Section~\ref{ExpCond})
and the characteristics of the undulator radiation 
for the positron energies, the types and lengths of the crystals, 
and the photon energies which will be available
in the experiments 
planned to be carried out within the PECU project\cite{PECU}.  

\begin{figure}[ht]
\begin{center}
\includegraphics*[scale=0.6]{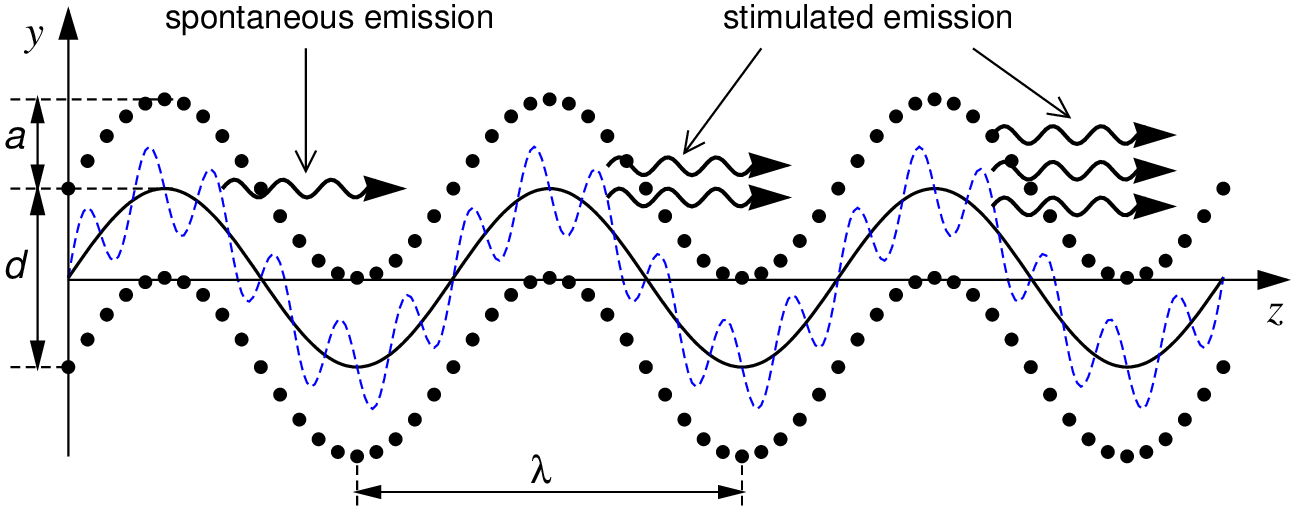}
\end{center}
\caption{Schematic representation of a crystalline undulator.
Circles denote the atoms belonging to 
neighboring crystallographic planes (separated by the distance $d$)
which are periodically bent.
The wavy line represents the trajectory of a positron 
which channels in between two planes, 
The profile of periodic bending is given by
$y(z)=a\sin(2\pi z/\lambda)$, where
the period $\lambda$ and amplitude $a$ satisfy the condition
$\lambda \gg a > d$. Other conditions - see eqs.~(\ref{AllConditions.5}).
}
\label{figure1.fig}
\end{figure}

A periodically bent crystal together with a bunch of ultra-relativistic 
charged particles which undergo planar channeling constitute a 
crystalline undulator, see Fig.~\ref{figure1.fig}.
In such a system there appears, in addition to the well-known channeling 
radiation, the  undulator type radiation 
which is due to the periodic motion of channeling particles which follow 
the bending of the crystallographic planes\cite{KSG1998,KSG1999}. 
The intensity and characteristic frequencies of this 
radiation can be varied by changing the beam energy 
and the parameters of the bending.
In the cited papers as well as in subsequent publications
(see the review Ref.~\citenum{KSG2004_review} and the references therein)
we proved a feasibility to create a short-wave crystalline undulator 
that will emit high-intensity, highly monochromatic radiation when pulses of 
ultra-relativistic 
positrons are passed through its channels\cite{ElectronUndul}. 
A number of corresponding novel numerical results were presented to
illustrate the developed theory, including, in particular, the
calculation of the spectral and angular characteristics of the new
type of radiation.
Later the importance of the novel concept of a crystalline undulator 
has been realized by other authors.

The scheme presented in Fig.~\ref{figure1.fig} leads also to the 
possibility of generating a stimulated emission of a free-electron 
laser type. 
The estimates carried out in Refs.~\citenum{KSG1999,KSG2004_review,SPIE2} 
showed that it is feasible to consider emission stimulation within 
the range of photon energies $10\dots 10^4$ keV (a Gamma-laser).
It was demonstrated\cite{SPIE1} also that 
the brilliance of radiation from a positron-based undulator 
in the energy range from hundreds of keV up to tens of MeV
is comparable to that of conventional light sources
(both existing and proposed) operating for much lower photon energies.

Once the feasibility of a positron-based crystalline undulator 
had been established theoretically,
it has become clear that further joint theoretical and
experimental efforts are needed to actually create this powerful 
source of radiation, the parameters of which can be easily 
tuned by varying the energy of the beam particles, by using different
crystals and by changing the parameters of the crystal bending.
All this constitute a complex, highly interdisciplinary, absolutely new 
and very promising field of research.
Constructing such a device is an extremely challenging task, 
to realize which it is necessary to bring together research groups from
various fields of expertise. 
Such a collaboration, created recently, has got the support from the 
European Commission within the framework of the PECU project\cite{PECU}.

The PECU project aims to develop in full the theory of of spontaneous 
and stimulated emission of radiation in crystalline undulator,
to use both theory and experiment to investigate the best methods and
materials for constructing the crystals, to carry out the experiments 
with the crystalline undulators.
The European researchers, involved in the project, 
are among the world leaders in various fields of research.
Theoretical support of the activities within the project 
will be carried out by researches from Frankfurt Institute for Advanced
Studies and Institut f\"{u}r Theoretische Physik (Univestit\"{a}t
Frankfurt am Main), whose pioneering works initiated the worldwide
activity in the field.
The key role in channeling experiments will be played by scientists
from the Aarhus University (Denmark), who have a long-term experience
in beam physics and studying the channeling phenomenon. 
Also, this institution has the necessary facilities to construct
periodically bent crystalline structures. 
The group from Mainz University (Germany) will carry out experiments 
on testing periodically bent structures. 
The group from Imperial College (the UK) will contribute 
its expertise in the free-electron laser physics.
The experiments with crystalline undulators will be performed at CERN and
INFN, where the positron beams, satisfying all the necessary criteria, 
are available.

For further referencing, let us mention the  
necessary conditions, which must be met in order to treat a 
crystalline undulator as a feasible scheme for devising 
a new source of electromagnetic radiation 
\cite{KSG1998,KSG1999,KSG2004_review}:
\begin{eqnarray}
\begin{cases}
C =4\pi^2\E a/  \dUmax \lambda^2 < 1
&   \mbox {- stable channeling},
\label{AllConditions.1}\\
d < a \ll \lambda
&   \mbox {- large-amplitude regime},
\label{AllConditions.2}\\
N = L/\lambda > 1
&    \mbox {- large number of periods},
\label{AllConditions.3}\\
L \lesssim \min\Bigl[\Ld(C),L_a(\om)\Bigr]
&    \mbox {- account for dechanneling and photon attenuation},
\label{AllConditions.4}\\
\Delta \E /\E \ll 1
&     \mbox {- low radiative losses}.
\label{AllConditions.5}
\end{cases}
\end{eqnarray}
Below we present a short description of the physics lying behind these 
conditions.

Stable channeling of an ultra-relativistic positron in a periodically 
bent channel 
is possible if the maximum centrifugal force $F_{\rm cf}$
is less than the maximal interplanar force $\dUmax$, i.e.
$C=F_{\rm cf}/\dUmax<1$. 
Expressing $F_{\rm cf}$ through the energy $\E$ of the projectile, the period 
and amplitude of the bending one formulates this condition 
as it is written in (\ref{AllConditions.5}).

The operation of a crystalline undulator should be considered in 
the  large-amplitude regime.
Omitting the discussion (see Ref.~\citenum{KSG1998,KSG1999,KSG2004_review}), 
we note that the limit 
$a/d>1$ accompanied by the condition $C\ll 1$  is mostly advantageous, 
since in this
case the characteristic frequencies of undulator and 
channeling radiation (see, e.g., Ref.~\citenum{Kumakhov2}) 
are well separated, so that the latter does not affect the parameters 
of the former,
whereas the intensity of undulator radiation becomes comparable or
higher than that of the channeling one\cite{KSG1998,KSG1999,KKSG00Tot}.
A strong inequality $a\ll\lambda$, 
resulting in elastic deformation of the crystal, 
leads to moderate values of the undulator parameter 
$p\equiv 2\pi \gamma a/\lambda \sim 1$ (here $\gamma=\E/mc^2$) 
which ensure that the emitted radiation is of the undulator type 
rather than of the synchrotron one\cite{Baier}.

The term 'undulator' implies that the number of periods
$N=L/\lambda$ is large (here $L$ denotes the crystal length).
In this case the emitted radiation bears the features of 
the undulator one, i.e.  
narrow, well-separated peaks, - harmonics,  
in spectral-angular distribution.
Therefore, the stronger the third inequality is the more pronounced
the features are.

The essential difference between  a crystalline undulator and a 
conventional one, 
based on the action of a magnetic (or electric) 
field\cite{RullhusenArtruDhez},
is that in the latter the beams of particles and photons move in vacuum 
whereas in the former \ -- \ in a crystalline medium, where
they are affected by {\em the dechanneling and the 
photon attenuation}. 
The dechanneling effect stands for a gradual increase in the 
transverse energy of a channeled particle due to inelastic collisions 
with the crystal constituents\cite{Lindhard}.
At some point the particle gains a transverse energy 
higher than the planar potential barrier and leaves the channel.
The average interval for a particle to penetrate
into a crystal until it dechannels is called the dechanneling length,
$\Ld$.
In a straight channel this quantity depends on the crystal, 
on the energy and the type of a projectile.
In a periodically bent channel there appears an additional 
dependence on the parameter $C$.
The intensity of the photon flux, which propagates through a crystal,
decreases due to the processes of absorption and scattering. 
The interval within which the intensity  decreases by a factor of $e$
is called the attenuation length,  $L_a(\om)$.
This quantity is tabulated for a number of elements 
and for a wide range of photon frequencies (see, e.g., 
Ref.~\citenum{ParticleDataGroup2006}).
The forth condition in (\ref{AllConditions.5})
takes into account severe limitation of the allowed values 
of the length $L$ of a crystalline undulator due to the dechanneling 
and the attenuation.

Finally, let us comment on the last condition in (\ref{AllConditions.5}).
For sufficiently large photon energies ($\om\gtrsim 10^2$ keV)
the restriction due to the attenuation becomes less 
severe than due to the dechanneling 
effect\cite{KSG1998,KSG1999,KSG2004_review}.
Then, $\Ld(C)$ introduces an upper limit on the 
length of a crystalline undulator.
Indeed, it was demonstrated\cite{SPIE1,Dechan01} that 
in the limit $L \gg \Ld$ the intensity of radiation is not 
defined by the expected number of undulator periods $L/\lambda$ but
rather is  formed in the undulator of the effective length $\sim\Ld$.
Since for an ultra-relativistic particle 
$\Ld \propto \E$\cite{BiryukovChesnokovKotovBook,Uggerhoj_RPM2005,Baier}, 
it seems natural that to increase the effective length 
one can consider higher energies. 
However, at this point another limitation manifests 
itself\cite{KSG1998,KSG1999,KSG00Loss}.
The coherence of an undulator radiation is only possible when
the energy loss $\Delta \E$ of the particle during its passage through the 
undulator is small, $\Delta \E\ll\E$.
This statement together with the fact, that for an ultra-relativistic 
projectile $\Delta \E$ is mainly due to the photon emission\cite{Baier}, 
leads to the 
conclusion that $L$ must be much smaller than the 
radiation length $\Lrad$, - 
the distance over which a particle converts its energy into radiation.

For a positron-based crystalline undulator 
a thorough analysis of the system  (\ref{AllConditions.5}) 
was carried out for the first time in 
Refs.~\citenum{KSG1998,KSG1999,SPIE1,KKSG00Tot,KSG00Loss,KSG2004_review}.
For a number of crystals the ranges of 
$\E$, $a$, $\lambda$ and $\om$ were established within which 
the operation of the crystalline undulator is possible.
These ranges include 
$\E=0.5\dots 10$ GeV, $a/d=10^1\dots10^2$,
$C=0.01\dots0.2$, 
$\om\gtrsim 10$ keV
and are common for all the investigated crystals.
The importance of exactly this regime of operation of the  positron-based
crystalline undulator was later realized by other 
authors\cite{BellucciEtal2003}.

\section{EXPERIMENTAL CONDITIONS} \label{ExpCond}

At the initial stage of the PECU project\cite{PECU}
two experiments on the measurement of the photon yield 
from positron-based crystalline undulators 
are planned to be carried out at CERN and INFN laboratories. 
Due to the experimental conditions and methods of preparations of
periodically bent crystalline structures
several parameters of the crystalline undulator are pre-set.
These parameters include\cite{Uggerhoj2006}:
\begin{itemize}
\item 
The positron beam energy is fixed at $\E=600$ MeV in 
the INFN experiment and at $\E=10$ GeV in the CERN one.

\item 
The crystalline undulators are to be produced by two methods.
The first method utilizes the technology of growing 
Si$_{1-x}$Ge$_x$ structures.
In this case, by varying the Ge content $x$ one can 
obtain periodically bent crystalline structure
\cite{MikkelsenUggerhoj2000,Darmstadt01}.
The technological restrictions imposed by this method on the
crystalline undulator length is that 
$L\leq 140\dots150\ \mu$m. 
\\
The periodic bending can also be achieved by
making regularly spaced grooves on the crystal surface\cite{BellucciEtal2003}. 
In this case, the crystalline planes in the vicinity of the defects become 
periodically bent. 
For the experiment within PECU a number of Si crystals prepared 
by laser-ablation method are available\cite{Uggerhoj2006,Connell2006}. 
The length of such crystalline undulators is $L=2$ or 4 mm, 
and the period of the structure is 
50, 100 and 200 $\mu$m.

\item 
A severe restriction on the emitted photon energy is anticipated 
in the INFN experiment where 
the experimental setup allows to register only $\hbar\om=20$ keV photons.

\end{itemize}

Theoretical support for these experiments implied to provide
an initial analysis of other parameters which can be varied in 
the crystalline undulators described above,
and, as a final result, to establish the ranges of parameters 
which lead to the 
highest yield of the undulator radiation.
Partly, the results of such analysis we present in sections
\ref{Undulator1} and \ref{Undulator2} 
for the following two cases (labeled below in the paper as 
'Undulator 1' and 'Undulator 2'):
\begin{eqnarray} 
\mbox{\it Undulator 1.}\ 
\mbox{The fixed parameters are:}
& & \E=0.6\ \mbox{GeV},\quad L=140\ \mu \mbox{m},\quad
 \hbar\om=20 \ \mbox{keV}.
\label{Undulator.1}\\
\mbox{\it Undulator 2.}\ 
\mbox{The fixed parameters are:}
& & \E=10\ \mbox{GeV},\quad L=150\ \mu \mbox{m}.
\label{Undulator.2}
\end{eqnarray}

\section{RESULTS OF CALCULATION} \label{Results}

\subsection{The formalism \label{Formalism}}
For the sake of reference let us outline the basic formulae
which we used in our calculations. 
A more detailed description of the formalism one can 
find in Refs.~\citenum{KSG1999,KSG2004_review,SPIE1,Dechan01}. 

The spectral distribution of the  energy $E$ of radiation
emitted in a crystalline undulator in the forward direction 
(i.e., $\theta =0$ with respect to the $z$ axis, see
Fig.~\ref{figure1.fig}) can be written in the following form\cite{SPIE1}:
\begin{eqnarray}
\left.
{\d^3 E \over \hbar\, \d\om\,\d\Om}\right|_{\theta=0} 
=
\alpha\gamma^2  
{ \eta^2 p^2 \over (1+p^2/2)^2 }\,  
\sin^2 {\eta\pi \over 2} 
\left[
 \bfJ_{{\eta + 1 \over 2}}\left(z\eta\right)
-
\bfJ_{{\eta - 1 \over 2}}\left(z\eta\right)
\right]^2
\cD_N(\eta),
\label{DechannelingAndAttenuation.1}
\end{eqnarray}
where $\alpha$ is the fine structure constant, $\gamma=\E/mc^2$,
$\bfJ_{\nu}(\xi)=\pi^{-1}\int_0^{\pi}\cos\Bigl(\nu\phi 
- \xi\sin\phi\Bigr)\, \d \phi$
is the Anger's function\cite{Gradshteyn},
$z=p^2/(4+2p^2)$ with $p=2\pi \gamma a/\lambda$ being the undulator parameter. 
The parameter $\eta$ is related to the frequency $\om$ of the emitted radiation
as follows
 \begin{eqnarray}
\om = {4 \gamma^2\omega_0 \over p^2 + 2}\, \eta,
\label{dspectral.4}
\end{eqnarray}
where $\om_0=2\pi c/\lambda$ is the undulator frequency.
The integer values of $\eta$, i.e. $\eta=n=1,2,3\dots$ define 
the  frequencies $\om_n$ of harmonics.
The factor $\sin^2 {\eta\pi/ 2}$ on the right-hand side of 
(\ref{DechannelingAndAttenuation.1}) ensures that only odd harmonics  
are emitted in the forward direction.
We also note that in this case the Anger's functions 
$\bfJ_{{(\eta \pm 1) / 2}}\left(z\eta\right)$ reduce to
the Bessel functions $J_{{(\eta \pm 1) / 2}}\left(nz\right)$ 
(e.g., Ref.~\citenum{Gradshteyn}), 
so that the expression in the square brackets acquires  
the form known in the theory of undulator radiation\cite{Baier,Alferov}.

A peculiar feature of the  undulator radiation,
which clearly distinguishes it from other types of electromagnetic radiation
by a charge moving in external fields, is that 
for each value of the emission angle (and, in particular, for $\theta=0$)
the spectral distribution consists of a set of narrow and equally spaced peaks
corresponding to different harmonics. 
In an ideal undulator (i.e., in which positrons and photons 
propagate in vacuum) the peak intensity is proportional to the 
squared number of periods.
Formally, it follows from the fact that $\d^3 E$ is proportional to 
$D_N(\eta)\equiv \Bigl(\sin (N\pi\eta)/ \sin(\pi\eta)  \Bigr)^2$ 
which behaves as
$N^2$ for $\eta =n$\cite{Baier,Alferov}.
This factor reflects the constructive interference of radiation emitted
from each of the undulator periods and is typical for any system which
contains $N$ coherent emitters.
Consequently, in an ideal undulator one can,  in principle, 
increase unrestrictedly the radiated intensity by 
increasing of the undulator length $L$ which, for a fixed $\lambda$, 
defines the number of undulator periods.

The situation is different for a crystalline undulator, where 
the number of channeling particles and the number of 
photons which can emerge from the crystal decrease with the growth of $L$.
In Ref.~\citenum{SPIE1,Dechan01} we analyzed quantitatively the influence 
of the dechanneling and the photon attenuation on the spectral-angular
 distribution.
The main result of these studies reads that the peak intensity of 
the radiation  is no longer proportional to $N^2$. 
Omitting the discussion, which can be found in the cited papers, 
we mention that 
in a crystalline undulator the factor $D_N(\eta)$ must be  substituted with 
$\cD_{N}(\eta)$, which depends not only on $N$ and $\eta$ but also on 
the ratios $\kp_{\rm d} ={L/\Ld(C)}$ and $\kp_{\rm a} ={L/ \La(\om)}$.
A convenient formula for $\cD_{N}(\eta)$ is as follows\cite{SPIE1}:
\begin{eqnarray}
\cD_{N}(\eta)
&=
\displaystyle{
{4N^2 \over \kp_{\rm a}^2 + 16N^2\sin^2\pi\teta}
\Biggl[
{\kp_{\rm a} \over \kp_{\rm a} -\kp_{\rm d}}\,\ee^{-\kp_{\rm d}}
-
{2\kp_{\rm d}-\kp_{\rm a}\over \kp_{\rm a} -\kp_{\rm d}}
{\kp_{\rm a}^2+4\phi^2\over(2\kp_{\rm d}-\kp_{\rm a})^2 +4\phi^2}\,
\ee^{-\kp_{\rm a}}
}
\nonumber\\
&\quad
\displaystyle{
-
2
\left(
\cos\phi
+
2\kp_{\rm d}\,
{
2\phi\,\sin\phi-(2\kp_{\rm d}-\kp_{\rm a})\cos\phi
\over
(2\kp_{\rm d}-\kp_{\rm a})^2 +4\phi^2}
\right)
\ee^{-(2\kp_{\rm d}+\kp_{\rm a})/2}
\Biggr],
}
\label{Spectr_Yes.2}
\end{eqnarray}
where  $\phi=2\pi\teta N$, $\teta = \eta-n$ and $n$ is a positive 
integer such as $n-1/2< \eta \leq n+1/2$.
Despite a cumbersome form of the right-hand side of 
(\ref{Spectr_Yes.2}) its main features can be easily understood. 
The most important is that, as in the case of an ideal undulator
(to which $\cD_{N}(\eta)$ reduces in the limit $\Ld(C) = \La(\om) = \infty$)
the main maxima of $\cD_{N}(\eta)$  correspond to the integer values of 
$\eta$, and, therefore, the harmonic frequencies are still defined by
(\ref{dspectral.4}) with $\eta=n$.
For finite $\Ld(C)$ and $\La(\om)$ the maximum value of   
$\cD_{N}(\eta)$ is smaller than $N^2$ whereas the width of the 
peak is larger than that in the corresponding ideal undulator.

To complete the descriptive part of the paper let us 
mention the method used to calculate the dechanneling and the attenuation
length.

The dechanneling length $L_{d}(C)$ in a periodically 
bent crystal is expressed via the dechanneling length 
$L_d(0)$ in the straight crystal as follows\cite{KSG1999,SPIE1}:
\begin{eqnarray} 
\Ld(C) = (1-C)^2\, \Ld(0),
\qquad
\Ld(0)
=
2\atf\, d\,{\E \over \Lambda} 
\, .
\label{DechannnelingAttenuation.3}
\end{eqnarray} 
Here  $d$ is the interplanar distance, 
$\atf$ is the Thomas-Fermi radius of the crystal atom.
The quantity $\Lambda= \ln{\sqrt{2\gamma}mc^2/I} -23/24$, with  
$I$ denoting the (average) ionization potential of the crystal atom,
is the Coulomb logarithm characterizing the ionization losses of an
ultra-relativistic positron in an amorphous medium. 
On the right-hand side of the second equation
it is implied that $\E$ is measured in GeV, $d$ and $\atf$ - in \AA, and
the result, $\Ld(0)$, - in cm.
The data on  $\Ld(0)$ for several straight channels are presented
in  Table~\ref{Ld.Table}. 
\begin{table}[h]
\caption{
Dechanneling length, $\Ld(0)$, in cm for $0.6$ and $10$ GeV 
positrons and for the (100), (110), (111) channels in 
straight Si and Ge crystals.
}
\label{Ld.Table}
\begin{center}       
\begin{tabular}{|r|rrr|rrr|} 
\hline
          & \multicolumn{3}{|c|}{$\E=0.6$ GeV} 
          & \multicolumn{3}{|c|}{$\E=10 $ GeV}   \\
Channel   & (100)  & (110)   & (111) &  (100)  & (110)   & (111)  \\
\hline
     Si   & 0.029  & 0.041   & 0.050 &  0.43   & 0.61    & 0.74   \\
     Ge   & 0.025  & 0.033   & 0.043 &  0.36   & 0.51    & 0.63   \\
\hline
\end{tabular}
\end{center}
\end{table} 

\begin{figure}[h]
\begin{center}
\includegraphics*[scale=0.45]{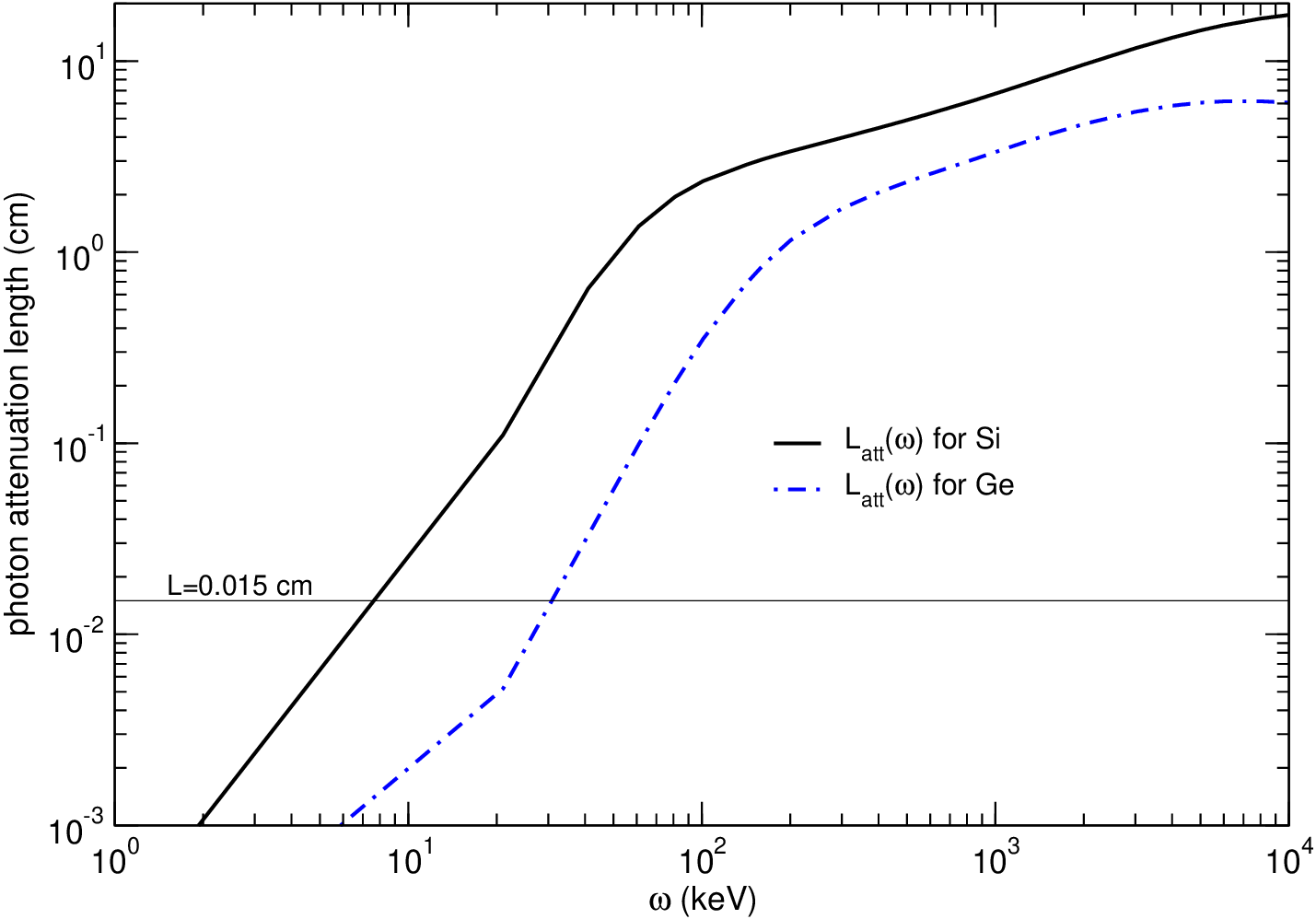}
\end{center}
\caption{
Photon attenuation lengths in Si and Ge.
The horizontal line indicates the length $L=0.015$ cm.
}
\label{Latt_SiGe.fig}
\end{figure}

The data on the attenuation lengths for various crystals can be found
in Ref.~\citenum{Hubbel}.
The dependence $\La(\om)$ for Si and Ge in a wide 
range of photon energies is presented in Fig.~\ref{Latt_SiGe.fig}.

Finally, we mention that 
the values of $\Ld\equiv \Ld(C)$ and $\La\equiv \La(\om)$  for 
a composite crystal Si$_{1-x}$Ge$_x$ can be estimated as follows:
\begin{eqnarray}
L_{\rm d,a} = (1-x)L_{\rm d,a}^{({\rm Si})} +x L_{\rm d,a}^{({\rm Ge})}.
\label{Frascati.3}
\end{eqnarray}                  
Since it is anticipated that the Ge content is small, namely 
$x\lesssim 0.01$, the values of $L_{\rm d,a}$ for the composite
are, practically, the same as for pure Si structure.

\subsection{Numerical results for 'Undulator 1'. 
\label{Undulator1}}

\begin{figure}[ht]
\begin{center}
\includegraphics[scale=0.7]{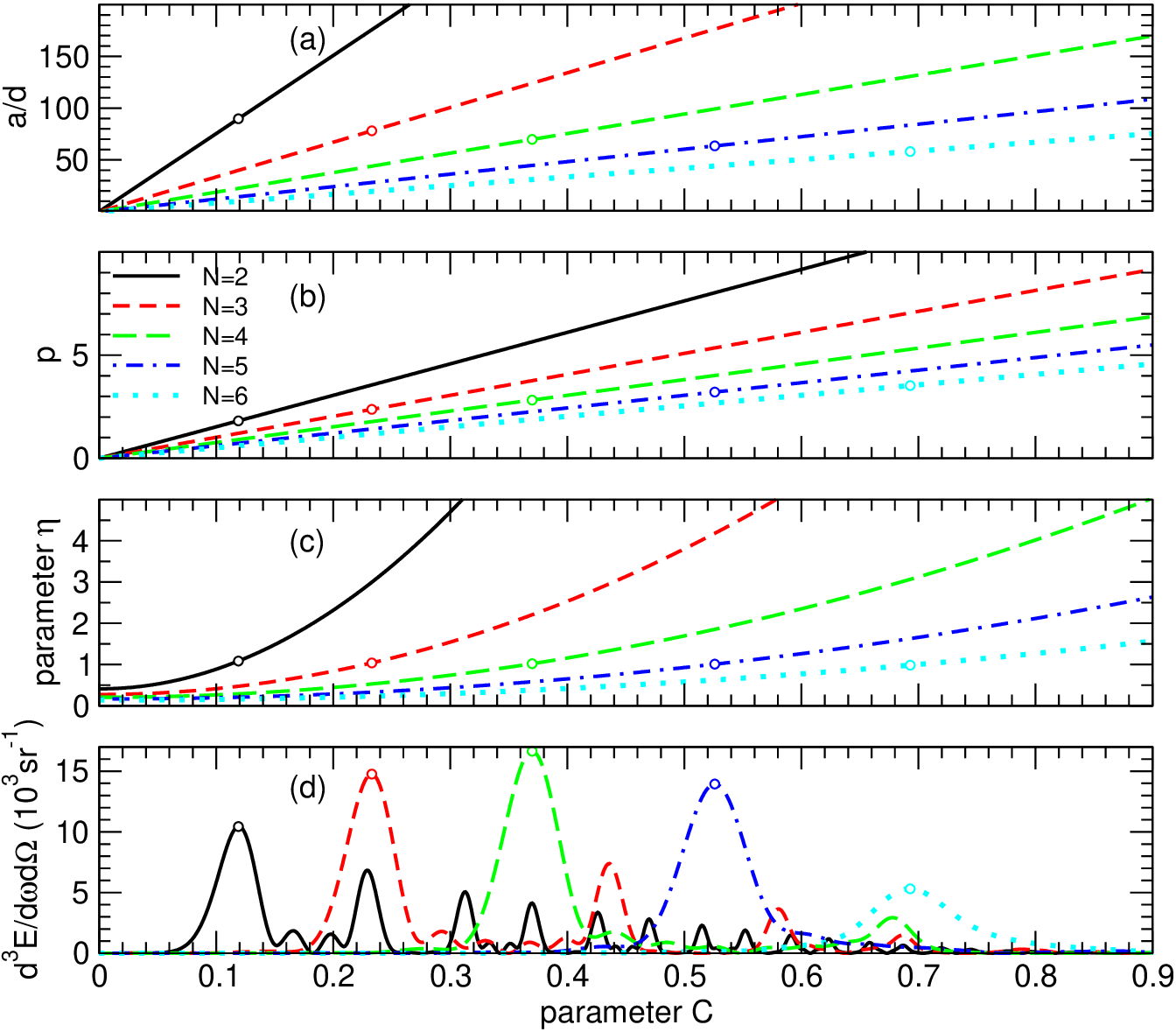}
\end{center}
\caption{
Parameters of the crystalline undulator  
(with the fixed parameters $\E=0.6$ GeV, $L=140\ \mu$m, 
$\hbar\om=20$ keV, see (\ref{Undulator.1}))
as functions of the parameter $C$  
and for various numbers of periods  $N$ as indicated.
\newline
Graphs (a) and (b) represent the dependences $a(C)/d$ and $p(C)$, - see
(\ref{Fixed_N_om.1}).
Graph (c) - represents $\eta(C)$, with  
$\eta$ defined in (\ref{dspectral.4}).
The dependence of $\d^3 E/\hbar \d \om \d\Om$ 
(see (\ref{DechannelingAndAttenuation.1})) on 
$C$ is presented in graph (d).
Open circles mark the parameters which correspond 
(for each $N$) to the main maxima of $\d^3 E/\d \om \d\Om$.   
}
\label{Und1_UndParam.fig}
\end{figure}

\begin{figure}[ht]
\begin{center}
\includegraphics[scale=0.55]{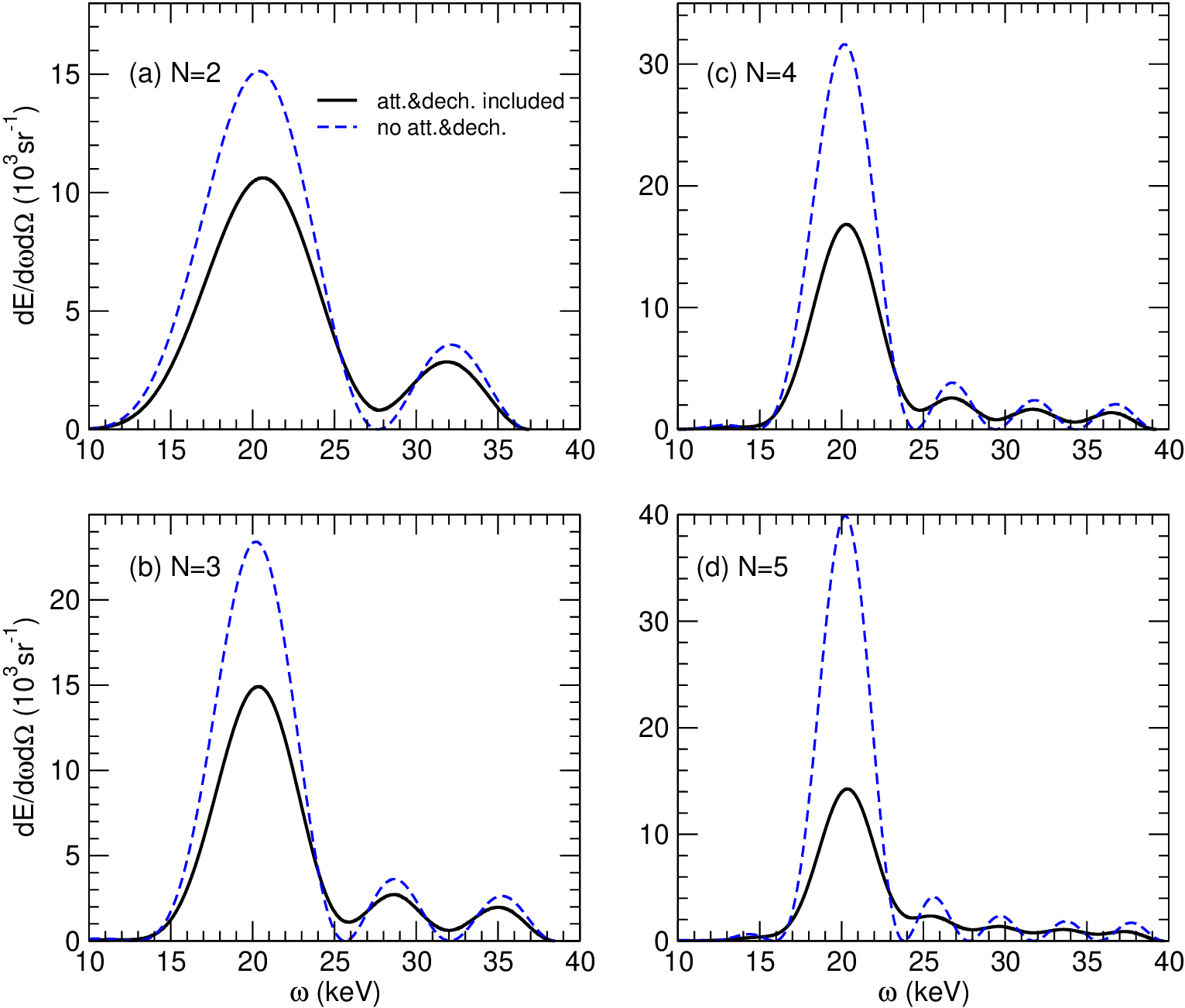}
\end{center}
\caption{
Spectral intensity of the undulator radiation 
in the region of the first harmonic maximum ($\hbar\om_1\approx 20$ keV)
calculated with the account for the dechanneling effect and 
photon attenuation (solid curves) and without (dashed curves). 
Four graphs correspond to the sets of parameters indicated in 
 Fig.~\ref{Und1_UndParam.fig} by open circles.
These parameters are:
\newline
(a) 
$N=2$, 
$C=0.12$, $a/d=90$, 
$p=1.8$, $\lambda=70$ $\mu$m;
\quad
(b) 
$N=3$, 
$C=0.23$, $a/d=78$, 
$p=2.4$, $\lambda=47$ $\mu$m;
\newline
(c) 
$N=4$, 
$C=0.37$, $a/d=70$, 
$p=2.8$, $\lambda=35$ $\mu$m;
\quad
(d) 
$N=5$, 
$C=0.53$, $a/d=64$, 
$p=3.2$, $\lambda=28$ $\mu$m.
}
\label{Und1_yes_no.fig}
\end{figure}

\begin{figure}[ht]
\begin{center}
\includegraphics[scale=0.55]{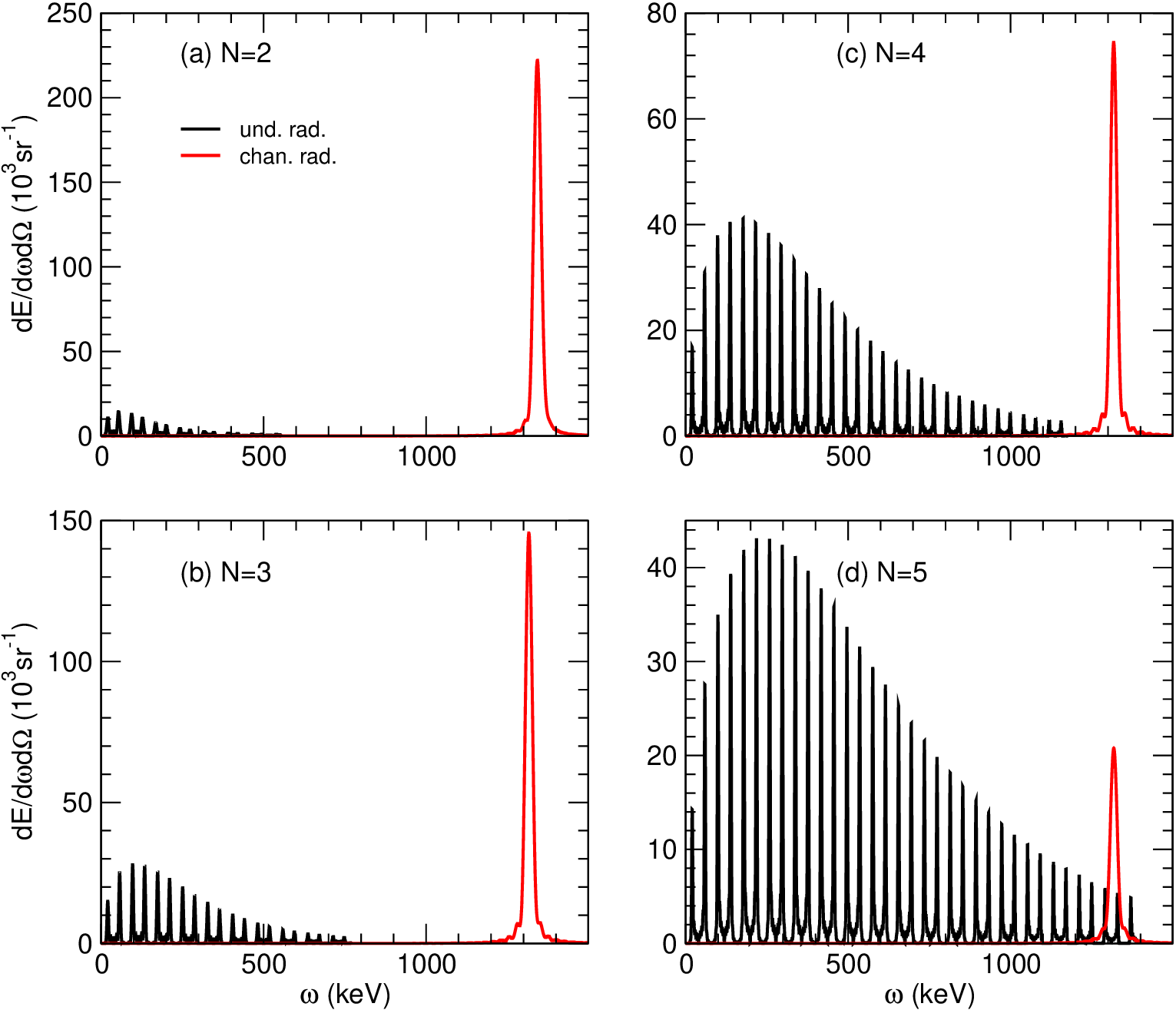}
\end{center}
\caption{
Comparison of the undulator (black curves) 
and the channeling (red curves) radiation spectra for  
the forward emission.
Each graph corresponds to the parameters which are indicated in the 
the caption to  Fig.~\ref{Und1_UndParam.fig}.
}
\label{Und1_UndChan_wide.fig}
\end{figure}

The goal of the quantitative analysis of the undulator 
with fixed values of $\E$, $L$ and $\om$  (see eq.~(\ref{Undulator.1})) 
was to establish the ranges of other parameters 
(these include, in particular, 
the number of periods $N$, the bending amplitude $a$, and the parameter $C$)
which ensure the largest photon yield.
The results of this analysis are presented in 
Figs.~\ref{Und1_UndParam.fig}-\ref{Und1_UndChan_wide.fig}.
The calculations were organized as follows.

As a first step, let us consider the undulators with different number 
of periods, $N=1,2,3\dots$.
The value of $N$ defines the period length: $\lambda_N=L/N$.
For fixed $\E$, $L$ and $N$ the first relation from
(\ref{AllConditions.5}) uniquely defines the dependence
of $a$ on $C$, and, as direct consequence, the dependence
of $p=2\pi\gamma a/\lambda$ on $C$:
\begin{eqnarray}
a(C) = {C\over N^2} \, \amax,
\qquad
p(C) = {C\over N}\, \pmax,
\label{Fixed_N_om.1}
\end{eqnarray}                  
where $\amax= \dUmax L^2/4\pi^2\E$ and $\pmax = \dUmax L/2\pi mc^2$ are the 
absolute maximum values of the amplitude and undulator parameter achieved
at $C=1$ and $N=1$.
The dependences $a(C)/d$ and $p(C)$ 
are presented in Figs.~\ref{Und1_UndParam.fig}(a)-(b). 
Fig.~\ref{Und1_UndParam.fig}(c) presents the dependence of the 
parameter $\eta$
on $C$ which, as it follows from (\ref{dspectral.4}), has the form 
$\eta(C)\propto 2 + p^2(C)$ and for each $N$ reaches its maximum value 
at $C=1$,  $\eta_{\max}=\eta(1)$.

Eq.~(\ref{Fixed_N_om.1}) and figs.~\ref{Und1_UndParam.fig}(a)-(c) 
show that generic type of each of the functions, -   
$a(C)$, $p(C)$ or $\eta(C)$, is independent on $N$, so that in each graphs 
the curves for different $N$ differ only quantitatively
(for fixed $C$ the larger values of $a$, $p$ and $\eta$ correspond 
to the smaller $N$'s).

These quantities, 
being used in (\ref{DechannelingAndAttenuation.1}),
allow one to analyze the dependence of 
the energy emitted at given frequency in the forward direction
$\d^3 E_N(C)\equiv \d^3 E/ \hbar\, \d\om\,\d\Om\Bigl|_{\theta=0}$ 
on $N$ and $C$, see Fig.~\ref{Und1_UndParam.fig}(d).
Let us mention several features of the functions $\d^3 E_N(C)$ and
relate them to other graphs presented in Fig.~\ref{Und1_UndParam.fig}.
Firstly, comparing the graphs (d) and (c) one notices that for each $N$ the 
(most pronounced) maxima  are located at those $C$-values which correspond 
to $\eta(C) =1,3,5\dots$, whereas
for even $\eta$ the spectrum $\d^3 E_N(C)=0$ in accordance
with general theory of the planar undulator radiation.
The maximum value of $\eta$ is much larger than one  for 
$N=1$ but $\eta_{\max} \approx 1.5$ in the case
$N=6$. 
As a result, the number of the maxima of $\d^3 E_N(C)$ 
decreases with $N$.
The open circles  in Fig.~\ref{Und1_UndParam.fig}(d) mark the
highest (for a given $N$) peak.
It is seen that the position $C_0$ of the highest peak  
gradually shifts towards larger values as $N$ increases.
This feature reflects the fact that  
for each $N$ the highest peak corresponds to $\eta = 1$,
which is achieved at larger $C$'s as $N$ grows, 
- Fig.~\ref{Und1_UndParam.fig}(c). 
The height of the peak exhibits a non-monotonous dependence on $N$.
One one understands this feature recalling eqs. 
(\ref{DechannelingAndAttenuation.1})-(\ref{DechannnelingAttenuation.3}).
From the first of these it follows that
$\d^3 E_N(C_0)\propto \cD_{N}(1)$ with the latter tending to increase 
with $N$, - 
note the factor $N^2$ on the r.h.s. of  (\ref{Spectr_Yes.2})).
However, as $C_0$ becomes larger the dechanneling lengths $\Ld(C_0)$, 
eq. (\ref{DechannnelingAttenuation.3}), decreases, so that 
the exponential factors $\exp(-\kp_{\rm d})\equiv \exp(-L/\Ld(C_0))$ 
reduce $\d^3 E_N(C_0)$.
Therefore, there exists a particular number of undulator periods 
which ensures the absolute maximum of the photon yield.
For the undulator with fixed values of $\E$, $L$ and $\om$  
(see eq.~(\ref{Undulator.1})) 
the absolute maximum is achieved for $N=4$ and  $C_0\approx 0.37$.
The corresponding values of the amplitude and undulator parameter are
$a(C_0)/d\approx 70$, $p(C_0)\approx 2.8$ 
(see Fig.~\ref{Und1_UndParam.fig}(a,b)). 

Figs.~\ref{Und1_yes_no.fig}(a)-(d) present spectral distribution
$\d^3 E/\hbar \d \om \d\Om$, - eq.~(\ref{Spectr_Yes.2}), as a function 
of photon energy 
within the interval including $\hbar \om = 20$ keV and calculated 
for different $N$ values.
For each $N$ the calculations were performed for the parameters 
$C, a, \eta$ and $p$ indicated in  Fig.~\ref{Und1_UndParam.fig} 
by open circles.
The solid curves were obtained with the dechanneling and photon 
attenuation effects
taken into account.
For the sake of comparison, the dashed curves represent the 
spectral distributions in the absence of these effects. 
The destructive role of these effects is clearly seen: 
the maxima of the solid curves are noticeably lower than 
those of the dashed curves.  

The data presented in Figs.~\ref{Und1_UndParam.fig} and \ref{Und1_yes_no.fig}
illustrate the procedure, following which one define the optimal parameters of 
the  crystalline undulator (with pre-set values of $\E$, $L$ and $\hbar\om$)
 in order to achieve the highest yield of the emission of the energy 20 keV.
However, the values of the undulator parameter $p$ in all four cases 
described in the
caption to Fig.~\ref{Und1_yes_no.fig} indicate that for $\hbar\om > 20$ keV  
one can expect higher values of the spectral intensities 
$\d^3 E/\hbar \d \om \d\Om$.
Indeed, as it follows from general theory of undulator radiation  
(see, e.g., \cite{Baier}), in the case  $p>1$ the number of emitted 
harmonics is $\sim p^3$ and the 
intensity of emission into the fundamental harmonic is not the highest one. 
To illustrate this statement in Fig.~\ref{Und1_UndChan_wide.fig} 
we present the results of calculation of $\d^3 E/\hbar \d \om \d\Om$ 
(eq.~(\ref{DechannelingAndAttenuation.1})) over the wide range of 
photon energies.
In each graph from this figure the characteristics of the undulator 
($C$, $p$, $a/d$ and $\lambda$) are as in the 
graph with the same $N$ from  Fig.~\ref{Und1_UndChan_wide.fig}.
The harmonic-like character of the spectral distribution manifests itself
in each graph, although it becomes more pronounced 
with the increase of $N$:
the number of the emitted harmonics (peaks) in 
Fig.~\ref{Und1_UndChan_wide.fig}(d) exceeds that seen in graph 
\ref{Und1_UndChan_wide.fig}(a) by a factor approximately 
equal to the cubed ratio of the corresponding undulator parameters 
($p=3.8$ and $p=1.8$, see Fig.~\ref{Und1_yes_no.fig}).
The figure demonstrates also that the undulators, initially 'tuned' to the 
emission of $\hbar\om=20$ keV, can be used to generate more energetic 
radiation and of a higher intensity.

The energy $\hbar\om_n = n\hbar\om_1$ of the harmonic of 
a sufficiently large order $n$ might become 
comparable with the characteristic energy of the channeling 
radiation\cite{Kumakhov2}.
Therefore, it is meaningful to compare the spectra of the undulator 
and the channeling 
radiation.
In Fig.~\ref{Und1_UndChan_wide.fig}
powerful and wide peaks in the region $\hbar\om \approx 1.3$ MeV
represents the spectral distribution 
$\d^3 E_{\ch} \equiv \d^3 E_{\ch}/\hbar \d \om \d\Om$
of the channeling radiation.
The latter was calculated using harmonic approximation for 
the interplanar potential.
Omitting the detailed discussion (see Ref.~\citenum{KSG00Loss})
let us briefly explain the decrease 
$\d^3 E_{\ch}$ with $N$.
Qualitatively, $\d^3 E_{\ch}$ is proportional to the (average) 
squared amplitude of the channeling oscillations, $a_{\ch}$.
In the straight channel $a_{\ch}\approx d/2$. 
In a periodically bent channel, the depth of the effective 
potential well decreases, leading to the decrease of the 
amplitudes of the channeling oscillations. 
Within the framework of the harmonic approximation 
$a_{\ch} \approx (1-C)d/2$\cite{BiryukovChesnokovKotovBook,KSG00Loss}. 
Therefore,  $\d^3 E_{\ch}\propto (1-C)^2$ - decreases with $C$.
Fig.~\ref{Und1_UndChan_wide.fig}(a) corresponds to the undulator with the 
smallest value of $C$,
whereas Fig.~\ref{Und1_UndChan_wide.fig}(d) - to the one with the largest $C$
(see  Fig.~\ref{Und1_UndParam.fig}(d) or/and 
the caption to Fig.~\ref{Und1_yes_no.fig}).
This explains the difference in the magnitudes of the channeling peaks.

\subsection{Numerical results for 'Undulator 2'.
\label{Undulator2}}
\begin{figure}[ht]
\begin{center}
\includegraphics[scale=0.7]{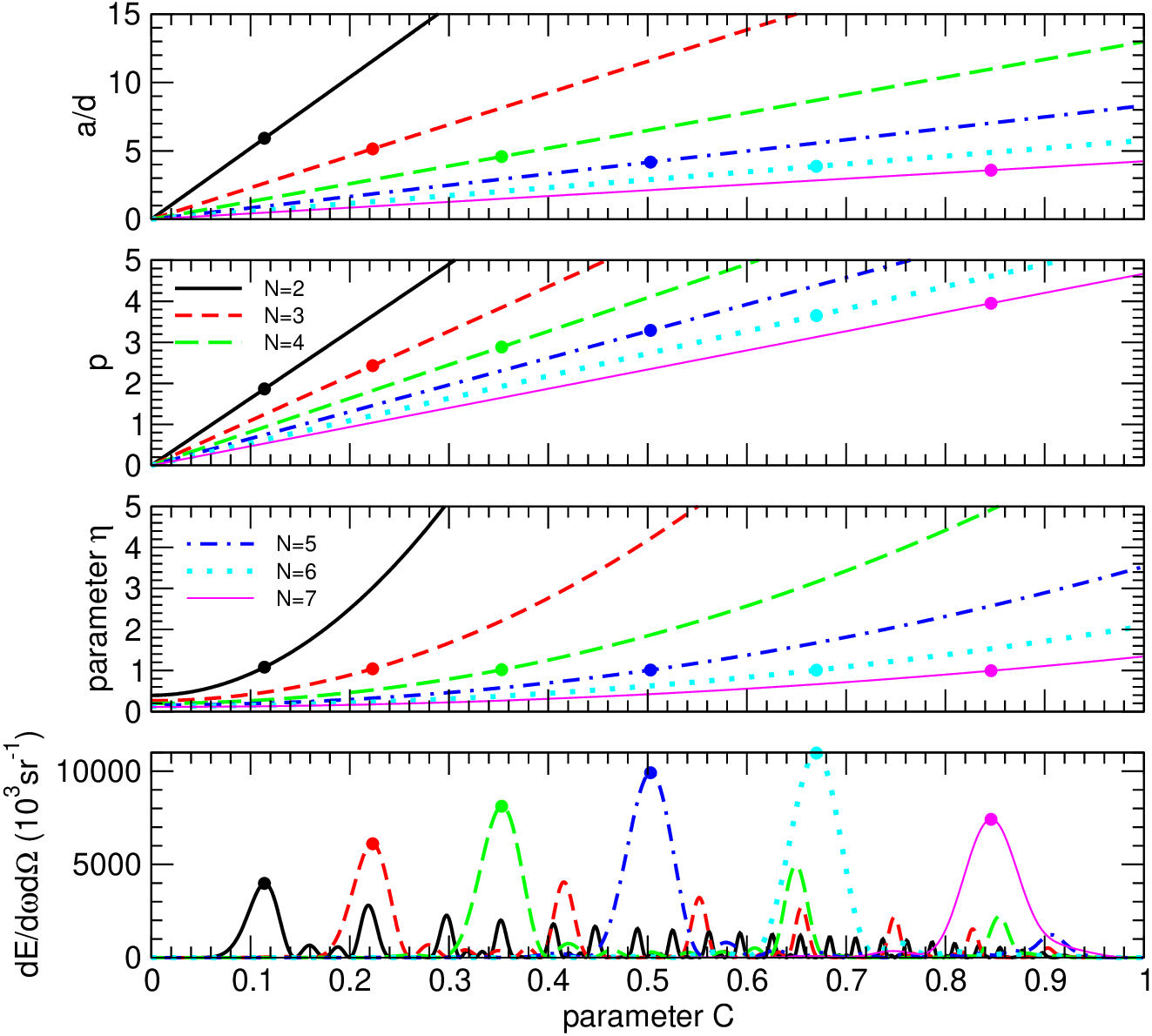}
\end{center}
\caption{
Parameters of the crystalline undulator  
(with the fixed parameters $\E=10$ GeV, $L=150\ \mu$m, see (\ref{Undulator.1}))
as functions of the parameter $C$  
and for various numbers of periods  $N$ as indicated.
The data refer to the undulator 'tuned' to 
the photon energy $\hbar\om = 5$ MeV.
\newline
Graphs (a) and (b) represent the dependences $a(C)/d$ and $p(C)$, - see
(\ref{Fixed_N_om.1}).
Graph (c) - represents $\eta(C)$, with  
$\eta$ defined in (\ref{dspectral.4}).
The dependence of $\d^3 E/\hbar \d \om \d\Om$ 
(see (\ref{DechannelingAndAttenuation.1})) on 
$C$ is presented in graph (d).
Circles mark the parameters which correspond 
(for each $N$) to the main maxima of $\d^3 E/\d \om \d\Om$.   
}
\label{Und2_UndParam.fig}
\end{figure}

\begin{figure}[ht]
\begin{center}
\includegraphics[scale=0.55]{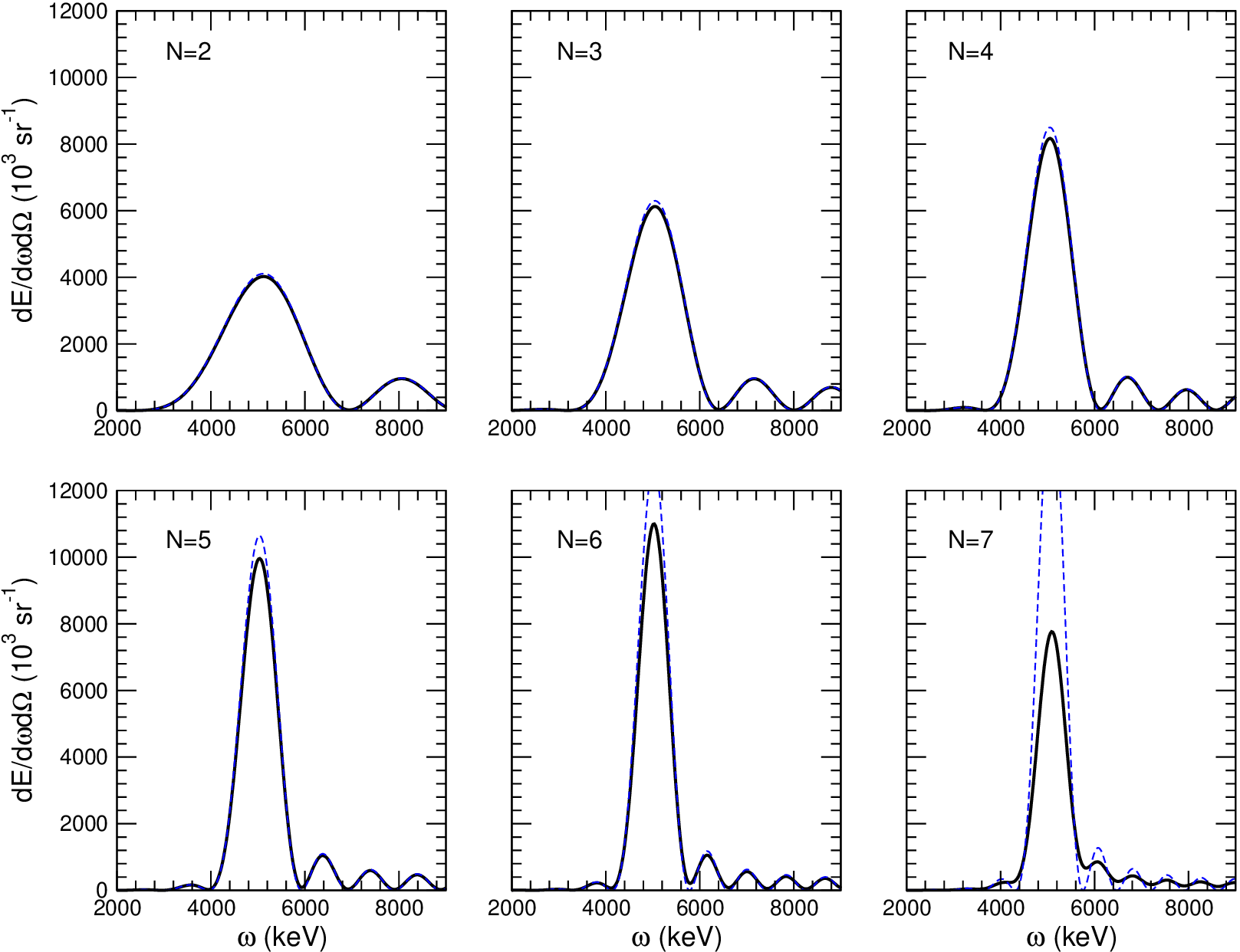}
\end{center}
\caption{
Spectral intensity of the undulator radiation 
in the region of the first harmonic maximum ($\hbar\om_1 = 5$ MeV)
calculated with (solid curves) and  without (dashed curves)
account for the dechanneling effect and photon attenuation.
Four graphs correspond to the sets of parameters indicated in 
 Fig.~\ref{Und2_UndParam.fig} by the circles.
These parameters are:
\newline
(a) 
$N=2$, 
$C=0.10$, $a/d=5.9$, 
$p=1.86$, $\lambda=75$ $\mu$m;
\quad 
(b) 
$N=3$, 
$C=0.20$, $a/d=5.1$, 
$p=2.4$, $\lambda=50$ $\mu$m;
\newline
(c) 
$N=4$, 
$C=0.35$, $a/d=4.6$, 
$p=2.90$, $\lambda=38$ $\mu$m;
\quad 
(d) 
$N=5$, 
$C=0.50$, $a/d=4.2$, 
$p=3.3$, $\lambda=30$ $\mu$m;
\newline
(e) 
$N=6$, 
$C=0.67$, $a/d=3.9$, 
$p=3.65$, $\lambda=25$ $\mu$m;
\quad
(f) 
$N=7$, 
$C=0.85$, $a/d=3.6$, 
$p=3.95$, $\lambda=21.4$ $\mu$m.
}
\label{Und2_yes_no.fig}
\end{figure}

\begin{figure}[ht]
\begin{center}
\includegraphics[scale=0.69]{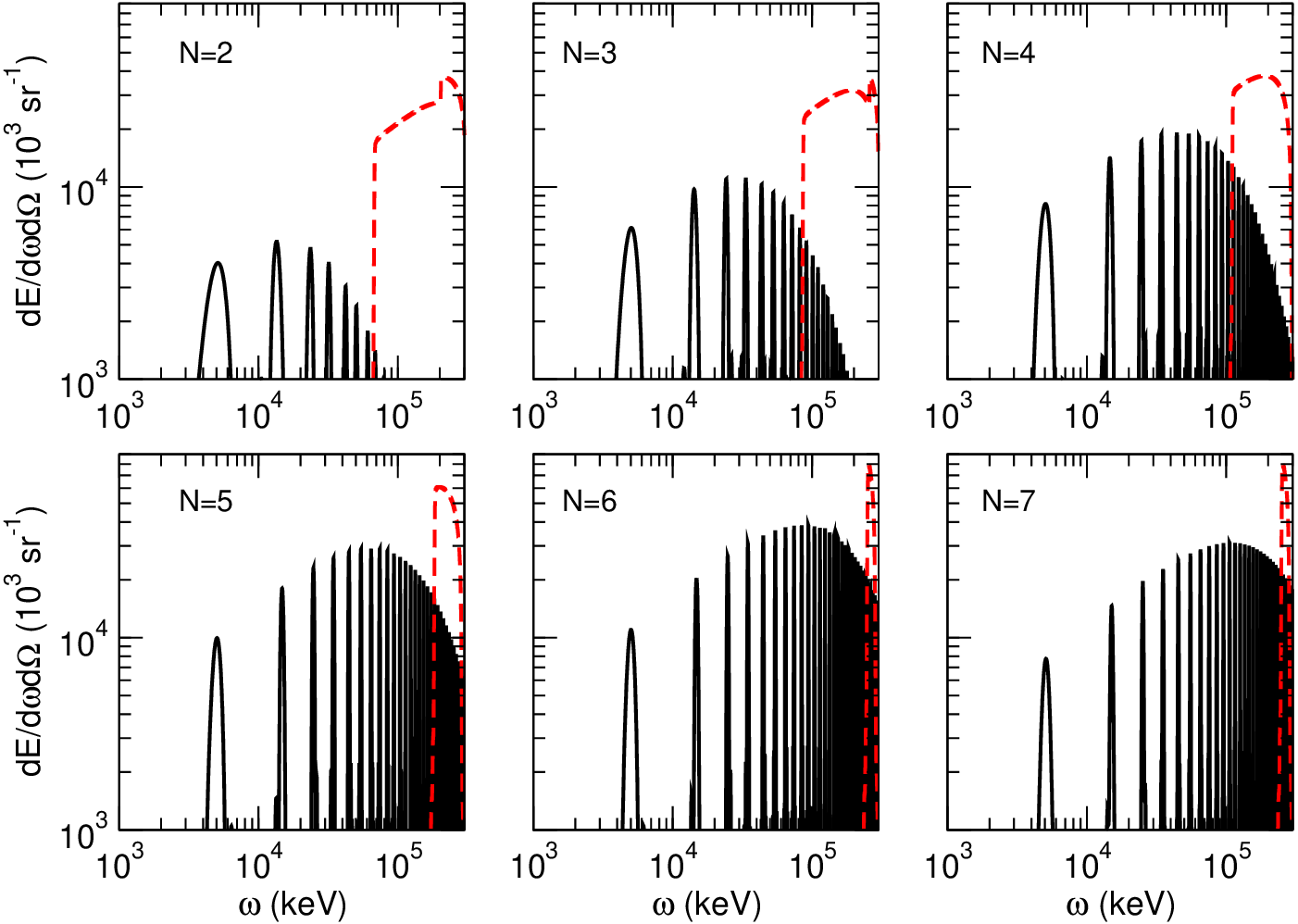}
\end{center}
\caption{
Spectral intensity of the undulator radiation
emitted in the forward direction calculated 
for $N=2\dots 7$ (as indicated) over the wide range of photon energies.
The parameters of undulators are indicated in 
the caption to  Fig.~\ref{Und2_yes_no.fig}.
The wide peaks (dashed curves) stand for the channeling radiation. 
}
\label{Und2_UndChan.fig}
\end{figure}

The results of numerical analysis of the second undulator, 
- eq.~(\ref{Undulator.2}), are presented in 
Figs.~\ref{Und2_UndParam.fig}-\ref{Und2_UndChan.fig}
The calculations were performed following, basically, the scheme 
outlined in Sec.~\ref{Undulator1}, though there were several specific
features.

Firstly, the energy of radiation was not limited by experimental
conditions. 
Therefore, it was meaningful to analyze the radiation within the 
range $\hbar\om = 10^2\dots 10^3$ keV where the photon attenuation
becomes much less pronounced, see Fig.~\ref{Latt_SiGe.fig}.
The data presented below refer to the undulator tuned to the first
harmonic energy $\hbar\om_1 = 5$ MeV. 
The corresponding attenuation length is nearly three orders of magnitude
larger than the crystal length $L=150\ \mu$m.
Therefore, the photon attenuation can be completely disregarded.
(In formal terms this means that one can put 
$\kp_{\rm a} ={L/ \La(\om)}=0$ on the right-hand side of 
(\ref{Spectr_Yes.2}).)

Secondly, the length of Undulator 2 only slightly 
exceeds that of Undulator 1, whereas the beam energy, $\E=10$ GeV, 
is nearly 20 times higher.
As a result, the dechanneling length $\Ld(0)$ becomes more that 
an order of magnitude larger than the crystal length 
(see Table~\ref{Ld.Table}). 
Therefore, one may expect that a strong inequality
$\Ld(C)=(1-C)^2\Ld(0) \gg L$ will be valid over a wide range of the 
bending parameter $C$ leading to a decrease of the influence
of the dechanneling effect on the photon yield.
Figs.~\ref{Und2_UndParam.fig}(a)-(d) present the dependences
$a(C)/d$, $p(C)$, $\eta(C)$ (see (\ref{Fixed_N_om.1})) and 
$\d^3 E_N(C)$ - the energy emitted in the forward direction,
calculated for several values of the number $N$ of undulator periods.
General features of all dependences are similar to those presented in 
Figs.~\ref{Und1_UndParam.fig}.
The qualitative differences are mostly pronounced for the $a(C)$ dependences:
it is seen that the $a/d$ values  in Fig.~\ref{Und2_UndParam.fig}(a)
are by more than an order of magnitude smaller than those in 
Fig.~\ref{Und1_UndParam.fig}(a).
This is solely due to the differences in the crystal length and the beam 
energy. 
Indeed, the factor $\amax= \dUmax L^2/4\pi^2\E$ (see(\ref{Fixed_N_om.1}))
for Undulator 2 is approximately 15 times less than for  Undulator 1.
On the contrary, the factor $\pmax = \dUmax L/2\pi mc^2$ is independent
on $\E$, and this results in close values of $p(C)$ for both undulators.
The peak intensities of $\d^3 E_N(C)$ in 
Fig.~\ref{Und2_UndParam.fig}(d) are three orders of magnitude 
higher than those in  Fig.~\ref{Und1_UndParam.fig}(d).
This increase is due to the following two reasons. 
Firstly, the factor $\gamma^2$ on the r.h.s. of
(\ref{DechannelingAndAttenuation.1}) ensures the increase by more than 
2 orders of magnitude.
The rest is donated by the enhancement of the factor $\cD_{N}(\eta)$ 
due to the decrease of the ratios $\kp_{\rm d} ={L/\Ld(C)}$ 
and $\kp_{\rm a} ={L/ \La(\om)}$.

Six graphs in Fig.~\ref{Und2_yes_no.fig} present spectral distribution
of the radiation in the forward direction as a function of photon energy 
within the interval including $\hbar \om = 5$ MeV and calculated 
for different $N$ values.
For each $N$ the calculations were performed for the parameters 
$C, a, \eta$ and $p$ marked in  Fig.~\ref{Und2_UndParam.fig} 
by the circles.
The solid and dashed curves were obtained with and without 
account for the positron dechanneling 
(the attenuation is negligibly small for a 5 MeV photon).
As mentioned above,  the influence of this effect in the case of 
Undulator 2 is much less pronounced that for Undulator 1
(see Fig.~\ref{Und1_yes_no.fig}).
Only in the two last graphs for $N=6$ and $N=7$ the account for the 
dechanneling results in a noticeable decrease of the photon yield.
These cases correspond (see the caption) to large values of the parameter
$C$ which, in turn, greatly reduces the dechanneling length 
(see eq.~(\ref{DechannnelingAttenuation.3})) making it comparable or even less
than the length of the crystal.

Finally, let us shortly comment on the graphs from Fig.~\ref{Und2_yes_no.fig}
which present the spectra $\d^3 E/\hbar \d \om \d\Om$ 
over the wide range of photon energies calculated for the undulators with 
the parameters enlisted in the caption to  Fig.~\ref{Und2_yes_no.fig}.
Note the double log scaled used in this figure in contrast to its analogue
for Undulator 1, - Fig.~\ref{Und2_yes_no.fig}.
Similar to the latter case the undulator radiation contains several
peaks corresponding to different harmonics. 
The energies $\hbar\om_n$ of the harmonics with low order $n$
are well-below the regions where the channeling radiation 
(the dashed curves) dominates.

\section{CONCLUSION} \label{Conclusion}

Theoretical investigations of the last 
decade\cite{KSG1998,KSG1999,KSG2004_review,SPIE1} have
proven that it is entirely realistic to use a positron-based
crystalline undulator for generating spontaneous radiation in 
a wide range of photon energies.
The parameters of such an undulator, being subject
to the restrictions mentioned in Sect.~\ref{Introduction}, can be 
tuned by varying the parameters of the bending, the positron energy 
and by choosing different crystals and applying different methods
to create periodically bent crystalline structures.

The efforts of the last years have succeeded in constructing 
the consortium\cite{PECU}, 
consisting of leading European groups, which will carry out
further theoretical and experimental studies of this phenomenon.
The latter, provided being successful, will become   
a very important step toward actual construction of
a new source of electromagnetic radiation at very high energies.

\acknowledgments    

We are grateful for Ulrik Uggerh{\o}j for providing the
data on the crystals length, ranges of photon energies and
energies of positron beams which are to be used in the experiments
in CERN and Frascati.
\\
This work has been supported by the European Commission 
(the PECU project, Contract No. 4916 (NEST)).



\begin{thebibliography}{99}
\bibitem{PECU}
        http://ec.europa.eu/research/fp6/nest/pdf
\bibitem{KSG1998}
         A.V.~Korol, A.V.~Solov'yov, W.~Greiner, 
         {\it J.~Phys.~G} {\bf 24}, L45 (1998).
\bibitem{KSG1999}
         A.V.~Korol, A.V.~Solov'yov, W.~Greiner, 
          {\it Int.~J.~Mod.~Phys. E} {\bf 8}, 49 (1999).
\bibitem{KSG2004_review}
         A.V.~Korol, A.V.~Solov'yov,  W.~Greiner, 
          {\it Int.~J.~Mod.~Phys. E} {\bf 13}, 867 (2004).
\bibitem{ElectronUndul}
        The feasibility of an {\it electron}-based crystalline 
        undulator was proven recently by 
        M.~Tabrizi, A.V.~Korol, A.V.~Solov'yov, W.~Greiner,
        submitted to {\it Phys. Rev. Lett.} (2006); 
        (arXiv: physics/0611012).
\bibitem{SPIE1}
        A.V.~Korol, A.V.~Solov'yov,  W.~Greiner, 
        Proc. SPIE - Int. Soc. Opt. Eng.  {\bf 5974}, Article 597405 (2005).
\bibitem{SPIE2}
        A.V.~Korol, A.V.~Solov'yov,  W.~Greiner, 
        Proc. SPIE - Int. Soc. Opt. Eng.  {\bf 5974}, Article 59740O (2005).

\bibitem{Kumakhov2}
         M.A.~Kumakhov, F.F.~Komarov,
         {\it Radiation From Charged Particles in Solids}
         (AIP, New York, 1989). 
\bibitem{KKSG00Tot}  
         W.~Krause, A.V.~Korol, A.V.~Solov'yov,  W.~Greiner, 
         {\it J.~Phys.~G} {\bf 26}, L87 (2000).
\bibitem{Baier}
         V.N.~Baier, V.M.~Katkov, V.M.~Strakhovenko, 
         {\it High Energy Electromagnetic Processes 
         in Oriented Single Crystals}
         (World Scientific,1998).
\bibitem{RullhusenArtruDhez}
         P.~Rullhusen, X.~Artru,  P.~Dhez,
         {\it Novel Radiation Sources using 
         Relativistic Electrons}
         (World Scientific, 1998).
\bibitem{Lindhard}
         J.~Lindhard, {\it Kong.~Danske~Vid.~Selsk.~Mat.-Fys.~Medd.}
        {\bf 34}, 14 (1965).
\bibitem{ParticleDataGroup2006}
         W.-M.~Yao {\it et al.},  {\it J.~Phys.~G} {\bf 33}, 1 (2006).
\bibitem{Dechan01} 
         A.V.~Korol, A.V.~Solov'yov,  W.~Greiner, 
         {\it J.~Phys.~G} {\bf 27}, 95 (2001).

\bibitem{BiryukovChesnokovKotovBook}
         V.M.~Biruykov,  Yu.A.~Chesnokov, V.I.~Kotov,
        {\it Crystal Channeling and its Application at 
         High-Energy  Accelerators}
         (Springer, Berlin, 1996).
\bibitem{Uggerhoj_RPM2005} 
         U.I.~Uggerh{\o}j, {\it Rev.~Mod.~Phys.} {\bf 77}, 1131 (2005).
\bibitem{KSG00Loss} 
         A.V.~Korol, A.V.~Solov'yov,  W.~Greiner, 
         {\it Int.~J.~Mod.~Phys.} E {\bf 9}, 77 (2000).

\bibitem{BellucciEtal2003}
         S.~Bellucci, S.~Bini, V.M.~Biryukov, Yu.A.~Chesnokov  et al, 
         {\it Phys.~Rev.~Lett.} {\bf 90}, 034801 (2003).
\bibitem{Uggerhoj2006}
         U.~Uggerh{\o}j, private communication (2006).
\bibitem{MikkelsenUggerhoj2000}  
         U.~Mikkelsen, E.~Uggerh{\o}j,
         {\it Nucl. Inst. and Meth. B}
         {\it Nuclear Instrum.  Methods B} {\bf 160}, 435 (2000).
\bibitem{Darmstadt01}
         A.~V.~Korol, W.~Krause, A.~V.~Solov'yov, W.~Greiner,
         {\it Nuclear Instrum.  Methods A} {\bf 483}, 
         455 (2002).
\bibitem{Connell2006}
         S.~H.~Connell, private communication (2006).

\bibitem{Gradshteyn}
         I.~S.~Gradshteyn, I.~M.~Ryzhik, 
         {\it Table of Integrals, Series and Products}
         (Academic Press, New York, 1965).
\bibitem{Alferov} 
         D.~F.~Alferov,  Yu.~A.~Bashmakov,   P.~A.~Cherenkov,
         {\it Sov.~Phys. - Uspekhi} {\bf 32}, 200 (1989).

\bibitem{Hubbel}
         J.~H.~Hubbel, S.~M.~Seltzer,
         {\it Tables of X-ray Mass Attenuation Coefficients},
          NISTIR 5632 - Web Version 1.02,
          http://physics.nist.gov/PhysRefData/XrayMassCoef/cover.html.

\end{thebibliography}
\end{document}